\documentclass[12pt]{article}

\newcommand{\eq}{\begin{equation}}
\newcommand{\eqx}{\end{equation}}
\newcommand{\eqn}{\begin{eqnarray}}
\newcommand{\eqnx}{\end{eqnarray}}
\newcommand{\f}[2]{\frac{#1}{#2}}
\newcommand{\cor}[1]{\left\langle{#1}\right\rangle}

\newcommand{\sg}{\sigma}
\renewcommand{\th}{\theta}
\newcommand{\Dl}{\Delta}
\newcommand{\dl}{\delta}
\newcommand{\al}{\alpha}
\newcommand{\der}{\partial}
\newcommand{\dd}{{\cal D}}
\newcommand{\xbar}{\bar{x}}
\newcommand{\eps}{\epsilon}
\newcommand{\qqqq}{\quad\quad\quad\quad}
\newcommand{\lra}{\longrightarrow}
\newcommand{\qb}{\bar{q}}
\newcommand{\gbar}{\bar{g}}

\newcommand{\rr}[4]{#1, {\it #2 \/}{\bf #3} #4}

\begin{document}

\title{String Fluctuations, AdS/CFT and the Soft Pomeron Intercept}

\author{Romuald A. Janik\footnote{{\tt
e-mail:ufrjanik@jetta.if.uj.edu.pl}}\\ \\
Service de Physique Th\'{e}orique  CEA-Saclay \\ F-91191
Gif-sur-Yvette Cedex, France\\ and\\
M. Smoluchowski Institute of Physics, Jagellonian University\\ Reymonta
4, 30-059 Cracow, Poland\footnote{Address after 9 October 2000.}}

\maketitle

\abstract{We study high energy scattering amplitudes in a strongly
coupled (confining) gauge theory using the AdS/CFT
correspondence. The scattering described by a Wilson line/loop
correlation function was shown earlier to correspond to
minimal surfaces of the helicoid type, and gave amplitudes with unit
intercept and linear trajectory. In this paper we find the correction
to the intercept from quadratic fluctuations of the string worldsheet
around the helicoid. The relevant term comes from analytical
continuation of a L\"uscher like term. It is coupling-constant
independent and proportional to the number of effective transverse 
flat dimensions. The shift of the intercept, under our assumptions, 
is $n_\perp /96$, and for $n_\perp=7,8$ gives respectively $0.0729$, 
$0.083$. Incidentally we note that this is surprisingly close to the
observed value of $0.08$.} 

\vfill

\section{Introduction}

The understanding of soft scattering phenomenae in QCD at high
energies remains a formidable challenge due to their strongly
nonperturbative character. The dominant behaviour at high energies (in
the soft regime) is governed by the experimentally observed soft
pomeron \cite{pomeron} giving amplitudes behaving like
$s^{1.08}$ \cite{DL}. The universality of the intercept, its
smallness and the precise value remains a mystery.

The recent developments of the AdS/CFT
correspondence \cite{ma98} (for a review see \cite{ma99}) provide an
effective set of tools which can be used to study strong coupling
physics of gauge theories.
In \cite{us2} scattering amplitudes in a confining theory in the
strong coupling limit were linked, through an identification with
Wilson line/loop correlation functions, with minimal surfaces of the helicoid
type in euclidean space. The resulting expressions were then 
analytically continued to Minkowski signature. The analytical
structure of these expressions involved branch cuts which, when taken
into account, gave contributions leading to inelastic amplitudes with linear
trajectories and unit intercept.

In this paper we would like to address the problem whether quantum
corrections might modify the intercept. 
We will evaluate quadratic fluctuations of the string
worldsheet around the classical solution obtained in \cite{us2}. We
will show that the term responsible for a shift of the intercept
arises by analytical continuation from a L\"uscher term. This property
is very specific to the helicoid minimal surface which appears when
calculating Wilson loop/line correlation functions related to
scattering amplitudes.

The plan of the paper is as follows. In section 2 we recall the
formalism for calculating scattering amplitudes from the AdS/CFT
correspondence used in \cite{us2,us1,Zahed}. In section 3 we
recapitulate some general features of the L\"uscher term arising from
string fluctuations. Then, in section 4, we recall, in a modified way
for the present purposes, the saddle point
results of \cite{us2} on the helicoid and scattering amplitudes 
and proceed, in section 5, to calculate the quantum fluctuations
around this minimal surface in euclidean space. We then perform
analytical continuation to Minkowski signature in order to obtain
corrections to the intercept.  
Finally we give our conclusions and list some remaining open problems.

\section{Scattering amplitudes}

In the eikonal approximation a scattering amplitude between two $q\qb$
pairs (resp. between $q$ and $\qb$) is expressed as a correlation
function of two Wilson loops (resp. lines) which follow the classical
straight line trajectories of the scattered particles
\cite{Nacht}-\cite{Nachtr}:
\eq
\cor{W(C_1)W(C_2)} \equiv \tilde{A}(s,L)
\eqx
where $s$ is the energy, $L$ is the impact parameter and
$\tilde{A}(s,L)$ is related to the momentum space amplitude through
\eq
\label{e.four}
\f{1}{s} A(s,t)=\f{i}{2\pi}\int d^2l \;e^{iq\cdot l}\;\tilde
A(s,L=|l|) \ .
\eqx
The loops (lines) $C_1$, $C_2$ are closed at infinity in order to
ensure gauge invariance.  
In the case of Wilson lines one has to introduce a finite temporal
length $T$ which acts as an IR regulator.

The above correlation function may be equivalently evaluated in
euclidean space, with the lines (loops) forming an angle $\th$ and
then performing the analytical continuation \cite{Megg}:
\eqn
\label{e.acont}
\theta &\lra& -i\chi \sim -i\log \f{s}{m^2} \ ,\nonumber\\
T &\lra& iT \ .
\eqnx

According to the AdS/CFT correspondence the correlation function is
expressed as the partition function of a string stretching between the
loops $C_1$ and $C_2$ \cite{loopsads}. In the strong coupling limit the saddle
point evaluation gives
\eq
\label{e.strans}
\cor{W(C_1)W(C_2)} \propto e^{-\f{1}{2\pi \al'} A_{minimal}}
\eqx
where $A_{minimal}$ is the area of the minimal surface spanned between
$C_1$ and $C_2$. $\al'$ depends on the gauge coupling as
$\al'=1/\sqrt{2g^2_{YM} N}$ in units of the AdS radius. In a confining
theory, when the impact parameter is sufficiently large, the minimal
surface is concentrated in the quasi-flat geometry near the horizon.
The analysis of the scattering amplitudes in this approximation has
been done in \cite{us2}. For the case of $q$-$\qb$ scattering with an
IR cut-off $T$ the relevant minimal surface is a helicoid of length
$T$. In a confining theory which we are considering here such a cut-off
may arise naturally in a physical manner from e.g. the string breaking
mechanism. For $q\qb$-$q\qb$ scattering the relevant minimal surface
is a ``tube'' formed by a pair of helicoids of length
$T_{int}$. $T_{int}$ is determined by a variational equation and
tends to zero $T_{int} \to 0$ as $s \to \infty$ (see \cite{us2}).

In this paper we would like to evaluate quadratic fluctuations around
the helicoid which is the basic building block of the relevant
scattering amplitudes. As is well known these $\al'$ corrections give
factors which are independent of $\al'$ and hence of the gauge coupling
constant (recall that the leading order term (\ref{e.strans}) 
behaves like $1/\al'$).

\section{The L\"uscher term}

Before we discuss the case of helicoid let us recall the basic results
for a single rectangular Wilson loop of size $T\times R$ relevant for
the calculation of the static $q\qb$ potential. A flat space
calculation \cite{luscher}-\cite{arvis} gives 
\eq
\label{e.luscher}
\cor{W(T\times R)}=e^{-\f{1}{2\pi \al'} TR} e^{n_\perp
\f{\pi}{24} \f{T}{R}} \ .
\eqx
The coulombic correction to the linear confining potential is the
L\"uscher term \cite{luscher}. Its main features are independence of
the string tension $\al'$ (related in the AdS/CFT correspondence to
the gauge coupling) and a large degree of model independence
(universality).

The term $n_\perp$ appearing in (\ref{e.luscher}) is the effective
number of transverse 
{\em massless} bosonic degrees of freedom on the worldsheet. 
Possible massive fluctuations are exponentially supressed in $R$. 
For a bosonic string in flat
$D$ dimensions $n_\perp=D-2$. A superstring in flat 10D spacetime has
$n_\perp=8-8=0$ (since massless bosonic and fermionic fluctuations cancel
out).

The problem of the existence of the L\"uscher term was reconsidered
from the point of view of the AdS/CFT correspondence in
\cite{olesen}-\cite{tseytlin2}. We will be especially
interested in the version dual to a confining theory.
In \cite{kinar} it was argued that for a specific black hole (BH) background
all the fermions become massive due to interactions
with the nontrivial RR background and hence do not contribute to the
L\"uscher term. In addition one of the $8=10-2$ transverse bosonic
fluctuations also acquires a mass and $n_\perp$ is effectively 7. It
seems that these arguments are quite generic. 

\section{The helicoid}

In a confining BH background, for a sufficiently large impact
parameter, the boundary conditions are effectively transported up to
the flat geometry near the horizon and it is enough to find the
minimal surfaces using a flat metric (see \cite{us2} in this context,
and \cite{olesen,estimates} for a general discussion).

The minimal surface spanned between two lines separated by a distance
$L$ and inclined at a relative angle $\th$ is a helicoid, which is
parameterized by
\eqn
t&=&\tau' \cos p\sg\\
y&=&\tau' \sin p\sg\\
x&=&\sg
\eqnx
where $p=\th/L$ and $\tau'=-T/2\ldots T/2$, $\sg=-L/2\ldots L/2$. In fact it
turns out to be more convenient to use a different parameterization
with $\tau'$ replaced by $\tau$ through $\tau'=\f{1}{p}\sinh
p\tau$. In the $\tau$, $\sg$ coordinates on the worldsheet the induced metric
$h_{ab}=\partial_a X^\mu \partial_b X^\mu$ is conformally
flat:
\eq
h_{ab}=(\cosh^2 p\tau) \dl_{ab} \ .
\eqx
The inverse relation between $\tau'$ and $\tau$
\eq
\tau=\pm \f{1}{p}\log\left(p\tau'+\sqrt{1+p^2\tau'^2}\right)
\eqx
defines the new intervals on which the parameters $\sg$ and $\tau$ are
defined, namely $\sg=-a/2\ldots a/2$ and $\tau=-b/2 \ldots b/2$ with
\eqn
\label{e.a}
a&=&L\\
\label{e.b}
b&=&\f{2L}{\th}\log\left(\f{\th T}{L}+\sqrt{1+\f{\th^2
T^2}{L^2}}\right) \ .
\eqnx
The classical area of the helicoid, entering formula (\ref{e.strans}), is now
\eq
\label{e.area}
A_{minimal}=\int d\sg d\tau \cosh^2 p\tau =\f{ab}{2}+\f{a \sinh
(bp)}{2p} \ .
\eqx

The area (\ref{e.area})
when inserted into (\ref{e.strans}) gives, according to the AdS/CFT
prescription, the correlation function of Wilson lines in euclidean
gauge theory (with the IR cut-off set by $T$).

\subsection*{Analytical continuation --- saddle point}

Following \cite{us2} let us now perfom, staying {\em within} gauge
theory, the analytical continuation
(\ref{e.acont}) of the area formula (\ref{e.area}). 
Due to the nonperturbative
character of the calculation the euclidean correlation function
related to (\ref{e.area}) is not single valued as can be seen from the 
presence of the logarithm. When performing the analytical continuation to
Minkowski signature we may not {\em a priori} rule out the possibility
of moving onto a different Riemann sheet. As in \cite{us2} we will
analyze the physical consequences of such a possibility. In the
following we thus have to keep track of possible branch cut contributions. 
The sides of the rectangle thus become
\eqn
a=L &\lra& L \\
\label{e.bcont}
b=\f{2L}{\th}\log (\ldots)&\lra& \f{2L}{\chi}(2\pi n+i\log(\ldots))
\eqnx
where $n$ is an integer.
If we now neglect the $\log$ with respect to $2\pi$ (which can be done
if the helicoid forms a part of the tube corresponding to the
scattering of two $q\qb$ pairs \cite{us2} or if we {\em assume} that
some confinement generated IR cut-off for $q$-$\qb$ scattering is small
enough) $b$ becomes just $4\pi L/\chi$ (or an 
integer multiple thereof\footnote{The $n$ of this paper corresponds to
$2n$ of \cite{us2}. The parametrization assumed here singles
out even branch cut contributions of \cite{us2}. The meaning of this
still remains to be understood.}). Inserting it back into the area 
formula (\ref{e.area}) we get the classical contribution to the
scattering amplitude (\ref{e.strans})
\eq
\exp\left\{-\f{1}{\al'}n\f{L^2}{\chi}\right\} \ .
\eqx
A transformation into momentum space \cite{us2} leads to an amplitude
with unit intercept and linear trajectory.
A detailed discussion is presented in \cite{us2}.

\section{Quadratic fluctuations}

In calculating the quadratic fluctuations we will make the following
assumptions about the underlying string theory: 
\begin{itemize}
\item[i)] the string theory is critical, i.e. conformally invariant,
\item[ii)] the bosonic massless fluctuations are described (up to
quadratic order) by the Polyakov action 
\eq
\f{1}{2\pi\al'} \int d\sg d\tau \sqrt{g}{g^{ab}}\der_a X^\mu \der_b
X^\mu \ .
\eqx
\end{itemize}

We will now expand the action around the classical solution
corresponding to the helicoid
\eq
X^\mu =\xbar^\mu+x^\mu \ .
\eqx
Since we assumed that the theory is critical, the background value of
the metric $\gbar_{ab}$ can be fixed to be any scalar multiple of the
induced metric $h_{ab}=\der_a \xbar \der_b \xbar$. It will be
convenient to take it to be flat:
\eq
\gbar_{ab}=\dl_{ab}=(\cosh p\tau)^{-2} h_{ab} \ .
\eqx
We will perform the calculation in the conformal gauge\footnote{A
calculation in a modified `static' gauge where two of the coordinates
are used to fix diffeomorphism invariance is much more involved and
leads to logarithmic divergences proportional to $\int
\sqrt{h}R^{(2)}(h_{ab})$ similar to those discussed in
\cite{forste,tseytlin1,tseytlin2}, which are difficult to cancel explicitly.}
\eq
\label{e.conf}
g_{ab}=\rho^2 \gbar_{ab}=\rho^2\dl_{ab} \ .
\eqx
Gauge fixing leads to the standard Faddeev-Popov determinant
\eq
\Delta_{FP}=\int \dd\eps_\tau \dd\eps_\sg \exp\left\{ \f{1}{2}\int
d\sg d\tau \nabla_k \eps_i \nabla_k \eps_i 
\right\}
\eqx
with the boundary conditions for the diffeomorphism ghosts being
\eq
\label{e.bc1}
\eps_\tau=0 \qqqq \der_\tau \eps_\sg=0
\eqx
at the boundary $\tau=\pm b/2$ and
\eq
\label{e.bc2}
\eps_\sg=0 \qqqq \der_\sg \eps_\tau=0
\eqx
at the boundaries at $\sg=\pm a/2$. 
The Dirichlet boundary condition comes from the requirement that the
diffeomorphisms preserve the boundary, while the Neumann boundary
condition is due to fixing the normal direction (see \cite{alvarez}).
We note that the corresponding ghost
action and boundary conditions depend only on the background
$\gbar_{ab}$ metric which is flat and hence are exactly the same as
for the case of a rectangular $a\times b$ Wilson loop.
Therefore
\eq
\Dl_{FP}=
\det \Dl_1
\eqx
where the subscript in $\Dl_1$ denotes that some boundary conditions are
changed from Dirichlet to Neumann according to
(\ref{e.bc1})-(\ref{e.bc2}). This change does not, however, modify the
determinant.

We will now consider the contributions of the bosonic fields
$X^\mu$. In general, the boundary conditions on $X^\mu$ are Dirichlet
up to a diffeomorphism acting on the boundary (see a discussion in
\cite{alvarez}). 
The Polyakov action expanded to quadratic order gives
\eq
\f{1}{2\pi\al'}\int d\sg d\tau \left( \der_a \xbar^\mu \der_a \xbar^\mu
+2\der_a \xbar^\mu \der_a x^\mu +\der_a x^\mu \der_a x^\mu \right) \ .
\eqx
As we have already fixed the diffeomorphism symmetry we may impose true
Dirichlet boundary conditions on all the fields \cite{alvarez}. 
The first term gives
just the area (\ref{e.area}), the second term then
can be seen to vanish, while the path integral over the $x^\mu$
gives
\eq
(\det \Dl)^{-\f{D}{2}} \ .
\eqx
To this we have to add the FP determinant. For a rectangle with the
flat metric $\dl_{ab}$ on the worldsheet $\Dl_{FP}$ just cancels two bosonic
modes and one finally gets
\eq
(\det \Dl)^{-\f{D-2}{2}} \ .
\eqx
Using $\zeta$-function regularization this gives
\eq
\label{e.zetdet}
\exp \left\{ \f{D-2}{2} \zeta_{rec}'(0) \right\}
\eqx
where $\zeta_{rec}'(0)$ is given by the expression (see e.g. \cite{book})
\eq
\label{e.zrec}
\zeta_{rec}'(0)=\f{1}{2}\log(2b)+\f{\pi}{12} \f{a}{b} +
2\sqrt{\f{a}{b}} \sum_{m,n=1}^\infty \sqrt{\f{n}{m}} K_{-1/2}
\left(2\pi m n \f{a}{b} \right) \ .
\eqx
In the limit $a \gg b$, the second term is dominant and for the
standard rectangular Wilson loop gives directly the L\"uscher
coulombic potential. Indeed denoting $n_\perp=D-2$ the dominant
expression is
\eq
\label{e.detfin}
\exp\left\{n_\perp \f{\pi}{24} \f{a}{b}\right\}
\eqx
with $a$, $b$ given by (\ref{e.a}) and (\ref{e.b}) respectively.

\subsection*{Analytical continuation --- fluctuations}

We will now perform the analytical continuation (\ref{e.acont}) of the 
fluctuation determinants (\ref{e.zetdet}), (\ref{e.detfin}). 
With the same assumptions
as in section 4 we have after continuation $a=L$ and $b=4\pi L/\chi$
(here we set the integer $n=1$).
For high energies $a\gg b$ and the dominant term in
(\ref{e.zrec}) will be just the `coulombic' L\"uscher term which gives
for each massless bosonic degree of freedom
\eq
\f{1}{2}\f{\pi}{12} \f{a}{b}=\f{\pi}{24} \f{\chi}{4\pi}=\f{1}{96}\log
s \ .
\eqx
Therefore the shift of the intercept is given by
\eq
\label{e.intercept}
s^{n_\perp \cdot \f{1}{96}} \ .
\eqx

Since the superstring moves in 10 dimensions, the generic number of
bosonic fluctuations is $n_\perp=10-2=8$ which would correspond to
$s^{8/96}=s^{0.08333}$. The arguments of \cite{kinar} suggest that one
of the bosonic modes becomes massive leading to
$s^{7/96}=s^{0.0729}$. The contribution of the massive bosonic and
fermionic modes will not be studied in this paper, but they should not
give rise to the `coulombic' $a/b$ terms responsible for the shift of
the intercept.

\section{Discussion}

In this paper we have studied corrections coming from fluctuations of
the string worldsheet to the scattering amplitudes obtained within the
AdS/CFT correspondence in \cite{us2}. The saddle point results of
\cite{us2} were obtained by evaluating correlation functions of Wilson
lines/loops in euclidean space as a function of the relative angle
$\th$ and performing analytical continuation to Minkowski space. The
possibility of analytic continuation to a different Riemann sheet
manifested itself in an inelastic amplitude with a linear trajectory.  

Here we concentrated on calculating the contribution of massless bosonic
modes of fluctuations around a helicoid and performed the same analytical
continuation, with the same choice of branch as in the saddle
point. The final result for a helicoid\footnote{Setting the integer
$n=1$.} is
\eq
\label{e.result}
(prefactors)\cdot s^{1+n_\perp \f{1}{96}+\f{\al'}{4}t}
\eqx
where the $(prefactors)$ can contain logarithms and weaker dependence
on $s$ (here we just concentrated on the dominant terms giving a
correction to the intercept). The intercept comes out to be $1.073$
for $n_\perp=7$ (favoured by AdS/CFT correspondence \cite{kinar}) and
$1.083$ for $n_\perp=10-2=8$.
In the above it was assumed that the helicoid was `short', i.e. that the
logarithm in (\ref{e.bcont}) could be neglected with respect to
$2\pi$. 

It is encouraging but indeed quite surprising that the result of the
analytical continuation for the `short' helicoid is so close to the
observed value of the soft pomeron intercept $s^{1.08}$ \cite{DL}.

The main result of this paper is the fact that a shift of the
intercept may appear in a natural way from analytical continuation of
the euclidean correlation function. This term comes directly from a
L\"uscher like term generated by string fluctuations and shares with
it {\em independence} of the coupling constant through $\al'$
\footnote{It is possible, in the AdS/CFT correspondence, that some coupling
constant dependence may be reintroduced through a modification of the
massless character of fluctuations of the string worldsheet. This is,
however, beyond the scope of this paper.}  . 
Moreover the numerical value (for the `short' helicoid) is quite small
and seems to be of the correct order of magnitude.

The above results suggest that the `short' helicoid geometry may be relevant
for the description of the soft pomeron, however one has to keep in mind
the possibility that other geometries may be more appropriate (like
the 'tube' geometry for $q\qb$ pair scattering) which may have similar
(but quantitatively different) fluctuation properties, thus pointing
towards a different `effective' number of degrees of freedom $n_\perp$.
Also, some ambiguities concerning the choice of a physically motivated branch 
remain to be clarified \cite{future}. 

Finally let us note that the string worldsheet which we considered in
this paper corresponding to a helicoid with a {\em finite} temporal
cut-off is quite far off from the standard picture of scattering of
asymptotic on-shell open string states. This might be the reason for
the possibility of deviation from the canonical open string intercept
of 1. Also the string calculation was done purely in euclidean space
(i.e. not in a direct scattering context). The shift of the intercept
arose when making the Wick rotation within gauge theory and performing an
analytical continuation to a different Riemann sheet.    

A number of open questions remain. Firstly, as emphasized above, a
careful analysis
of fluctuations around the `tube' geometry relevant for scattering of
$q\qb$ pairs remains to be done \cite{future}. It is exactly this case which
is physically much better defined, IR finite, and would be expected to
correspond to the soft pomeron rather than the single
helicoid. Therefore a lot of caution has to be maintained in
interpreting (\ref{e.result}). In addition possible terms with
weaker energy behaviour have not been determined so far.

Secondly it would be interesting to reproduce the above results in a
`static' gauge calculation and investigate in more detail the
subleading contributions of GS fermions and massive modes.

A more in depth understanding of the meaning of the branch cut
structure of the Wilson line/loop correlation functions would also be
very interesting. It is tempting to speculate that some features of
this structure may be more universal and stable with respect to
deformations or modifications of the minimal surfaces.

\subsubsection*{Acknowledgements}

I would like to thank Robi Peschanski for numerous discussions and comments.
This research was partially supported by KBN grants 2P03B08614 and
2P03B00814.

\end{document}